\documentstyle[preprint,pra,aps]{revtex}

\begin{document}

\bibliographystyle{prsty}
\draft
%\preprint{submitted to Phys.Rev. A}
\tighten

\title{Nonclassical correlations of photon number and 
field components in the vacuum state}

%Nonclassical correlations in the quantum jump backaction of
%quadrature component measurements performed on the photon vacuum}
%

\author{Holger F. Hofmann and Takayoshi Kobayashi}
\address{Department of Physics, Faculty of Science, University of Tokyo,\\
7-3-1 Hongo, Bunkyo-ku, Tokyo113-0033, Japan}

\author{Akira Furusawa}
\address{Nikon Corporation, R\&D Headquarters,\\
Nishi-Ohi, Shinagawa-ku, Tokyo 140-8601, Japan}

%\author{Holger F. Hofmann\\Department of Physics, Faculty of Science, 
%University of Tokyo\\7-3-1 Hongo, Bunkyo-ku, Tokyo113-0033, Japan}

\date{\today}

\maketitle

\begin{abstract}
It is shown that the quantum jumps in the photon number $\hat{n}$
from zero to one or more photons induced by backaction evasion quantum 
nondemolition measurements of a quadrature component $\hat{x}$ of the 
vacuum light field state are strongly correlated with the quadrature 
component measurement results.
This correlation corresponds to the operator expectation value
$\langle \hat{x}\hat{n}\hat{x}\rangle$ which is equal to one fourth for
the vacuum even though the photon number eigenvalue is zero.
Quantum nondemolition measurements of a quadrature component can
thus provide experimental evidence of the nonclassical operator 
ordering dependence of the correlations between photon number and field 
components in the vacuum state.
\end{abstract}

\section{Introduction}
One of the main differences between quantum mechanics and classical
physics is the impossibility of assigning well defined values to 
all physical variables describing a system. As a consequence, all
quantum measurements necessarily introduce noise into the system.
A measurement which only introduces noise in those variables that 
do not commute with the measured variable is referred to as a 
quantum nondemolition (QND) measurement \cite{Cav80}. In most of the
theoretical and experimental investigations 
\cite{Lev86,Fri92,Bru90,Hol91,Yur85,Por89,Per94}, the focus has
been on the overall measurement resolution and on the reduction of 
fluctuations in the QND variable as observed in the correlation between 
the QND measurement results and a subsequent destructive measurement 
of the QND variable. However, at finite resolution, quantum nondemolition
measurements do not completely destroy the original coherence between 
eigenstates of the QND variable \cite{Imo85,Kit87}. By correlating the 
QND measurement result with subsequent destructive measurements of a 
noncommuting variable, it is therefore possible to determine details of 
the measurement induced decoherence \cite{Hof20}. 

In particular, QND measurements of a quadrature component of the light
field introduce not only noise in the conjugated quadrature component,
but also in the photon number of a state. By measuring a quadrature
component of the vacuum field, ``quantum jumps'' from zero photons
to one or more photons are induced in the observed field. It is shown in
the following that, even at low measurement resolutions, the 
``quantum jump'' events are strongly correlated with extremely high 
measurement results for the quadrature component. This correlation 
corresponds to a nonclassical relationship between the continuous
field components and the discrete photon number, which is directly
related to fundamental properties of the operator formalism. 
Thus, this experimentally observable correlation of photon number and 
fields reveals important details of the physical meaning of quantization.

In section \ref{sec:qnd}, QND measurements of a 
quadrature component $\hat{x}$ of the light field are discussed and
a general measurement operator $\hat{P}_{\delta\! x}(x_m)$ describing 
a minimum noise measurement at a resolution of $\delta\!x$ is derived.
In section \ref{sec:vac}, the measurement operator is applied to the 
vacuum field and the measurement statistics are determined.
In section \ref{sec:fundop}, the results are compared with fundamental 
properties of the operator formalism.
In section \ref{sec:ex}, an experimental realization of photon-field
coincidence measurements is proposed and possible difficulties are discussed.
In section \ref{sec:int}, the results are interpreted in the context
of quantum state tomography and implications for the interpretation 
of entanglement are pointed out.
In section \ref{sec:concl}, the results are summarized and conclusions
are presented.

%=========================================================

\section{QND measurement of a quadrature component}
\label{sec:qnd}

Optical QND measurements of the quadrature 
component $\hat{x}_S$ of a signal mode 
$\hat{a}_S = \hat{x}_S + i \hat{y}_S$
are realized by coupling the signal to a a meter mode 
$\hat{a}_M = \hat{x}_M + i \hat{y}_M$ in such a way that the 
quadrature component $\hat{x}_M$ of the meter mode is shifted
by an amount proportional to the measured signal variable
$\hat{x}_S$. This measurement interaction 
can be described by a unitary transformation operator, 
\begin{equation}
\hat{U}_{SM} = \exp\left(-i\; 2 f \hat{x}_S\hat{y}_M\right),
\end{equation}
which transforms the quadrature components of meter and signal to
\begin{eqnarray}
\label{eq:shift}
\hat{U}_{SM}^{-1}\;\hat{x}_S\;\hat{U}_{SM} &=& \hat{x}_S
\nonumber \\[0.2cm]
\hat{U}_{SM}^{-1}\;\hat{y}_S\;\hat{U}_{SM} &=& \hat{y}_S - f \hat{y}_M
\nonumber \\[0.2cm]
\hat{U}_{SM}^{-1}\;\hat{x}_M\;\hat{U}_{SM} &=& \hat{x}_M + f \hat{x}_S
\nonumber \\[0.2cm]
\hat{U}_{SM}^{-1}\;\hat{y}_M\;\hat{U}_{SM} &=& \hat{y}_M.
\end{eqnarray}
In general, the unitary measurement interaction operator 
$\hat{U}_{SM}$ creates entanglement between the signal and the meter
by correlating the values of the quadrature components.
Such an entanglement can be realized experimentally by squeezing
the two mode light field of signal and meter using optical
parametric amplifiers (OPAs) \cite{Yur85,Por89,Per94}. The measurement
setup is shown schematically in figure \ref{setup}. Note that the 
backaction changing $\hat{x}_S$ is avoided by adjusting the interference
between the two amplified beams. Therefore, the reflectivity of the beam
splitters depends on the amplification. A continuous adjustment of the 
coupling factor $f$ would require adjustments of both the pump beam 
intensities of the OPAs and the reflectivities of the beam splitter as
given in figure \ref{setup}.

If the input state of the meter is the vacuum field state, 
$\mid \mbox{vac.} \rangle$, and the signal field state is given by
$\mid \Phi_S \rangle$, then the entangled state created by the
measurement interaction is given by 
\begin{eqnarray}
\hat{U}_{SM}\mid \Phi_S; \mbox{vac.}\rangle &=&
\int d\!x_S d\!x_M\; \langle x_S \mid \Phi_S \rangle \;
             \langle x_M - f x_S \mid \mbox{vac.} \rangle
             \; \mid x_S; x_M \rangle
\nonumber \\
&=& \int d\!x_S d\!x_M\; \left(\frac{2}{\pi}\right)^{\frac{1}{4}}
\exp\left(-(x_M-f x_S)^2\right) \langle x_S \mid \Phi_S \rangle
 \; \mid x_S; x_M \rangle.
\end{eqnarray}
Reading out the meter variable $x_M$ removes the entanglement by
destroying the coherence between states with different $x_M$. 
It is then possible to define a measurement operator $\hat{P}_f(x_M)$
associated with a readout of $x_M$, which acts only on the initial
signal state $\mid \Phi_S \rangle$. This operator is given by
\begin{eqnarray}
\langle x_S \mid \hat{P}_f(x_M) \mid \Phi_S \rangle &=&
\langle x_S; x_M \mid\hat{U}_{SM}\mid \Phi_S; \mbox{vac.}\rangle 
\nonumber \\
&=&
\left(\frac{2}{\pi}\right)^{\frac{1}{4}}
\exp\left(-(x_M-f x_S)^2\right) \langle x_S \mid \Phi_S \rangle.
\end{eqnarray}
The measurement operator $\hat{P}_f(x_M)$ multiplies the probability
amplitudes of the $\hat{x}_S$ eigenstates with a Gaussian statistical 
weight factor given by the difference between the eigenvalue 
$x_S$ and the measurement result $x_M/f$. By defining
\begin{eqnarray}
x_m &=& \frac{1}{f} x_M
\nonumber \\
\delta\!x &=& \frac{1}{2f},
\end{eqnarray} 
the measurement readout can be scaled, so that the average results
correspond to the expectation value of $\hat{x}_S$. 
The normalized measurement operator then reads
\begin{equation}
\label{eq:project}
\hat{P}_{\delta\!x}(x_m) = \left(2 \pi \delta\!x^2\right)^{-1/4} 
\exp \left(-\frac{(x_m-\hat{x}_S)^2}{4\delta\!x^2}\right).
\end{equation}
This operator describes an ideal quantum nondemolition measurement 
of finite resolution $\delta\!x$. 
The probability distribution of 
the measurement results $x_m$ is given by
\begin{eqnarray}
\label{eq:prob}
P(x_m) &=& \langle \Phi_S \mid \hat{P}^2_{\delta\!x}(x_m) \mid \Phi_S \rangle
\nonumber \\
&=& \frac{1}{\sqrt{2\pi \delta\!x^2}}\int d\!x_S\;  
\exp\left(-\frac{(x_S-x_m)^2}{2\delta\!x^2}\right)
|\langle x_S \mid \Phi_S \rangle |^2  
. 
\end{eqnarray}
Thus the probability distribution of measurement results is equal to
the convolution of $|\langle x_S \mid \Phi_S \rangle |^2 $ with a Gaussian
of variance $\delta\! x$. The corresponding averages of $x_m$ and $x_m^2$
are given by
\begin{eqnarray}
\label{eq:av}
\int d\!x_S\; x_m P(x_m) 
        &=& \langle \Phi_S \mid \hat{x}_S \mid \Phi_S \rangle
\nonumber \\
\int d\!x_S\; x_m^2 P(x_m) 
        &=& \langle \Phi_S \mid \hat{x}_S^2 \mid \Phi_S \rangle
            + \delta\!x^2. 
\end{eqnarray}
The measurement readout $x_m$ therefore represents the actual value of
$\hat{x}_S$ within an error margin of $\pm \delta\!x$.
The signal state after the measurement is given by
\begin{equation}
\label{eq:state}
\mid \phi_S(x_m)\rangle = \frac{1}{\sqrt{P(x_m)}}
\hat{P}_{\delta\!x}(x_m) \mid \Phi_S \rangle. 
\end{equation}
Since the quantum coherence between the eigenstates of $\hat{x}_S$
is preserved, the system state is still a pure state after the
measurement. The system properties which do not commute with
$\hat{x}_S$ are changed by the modified statistical weight
of each eigenstate component. Thus the physical effect of noise in
the measurement interaction is correlated with the measurement
information obtained. 

%=========================================================

\section{Measurement of the vacuum field}
\label{sec:vac}

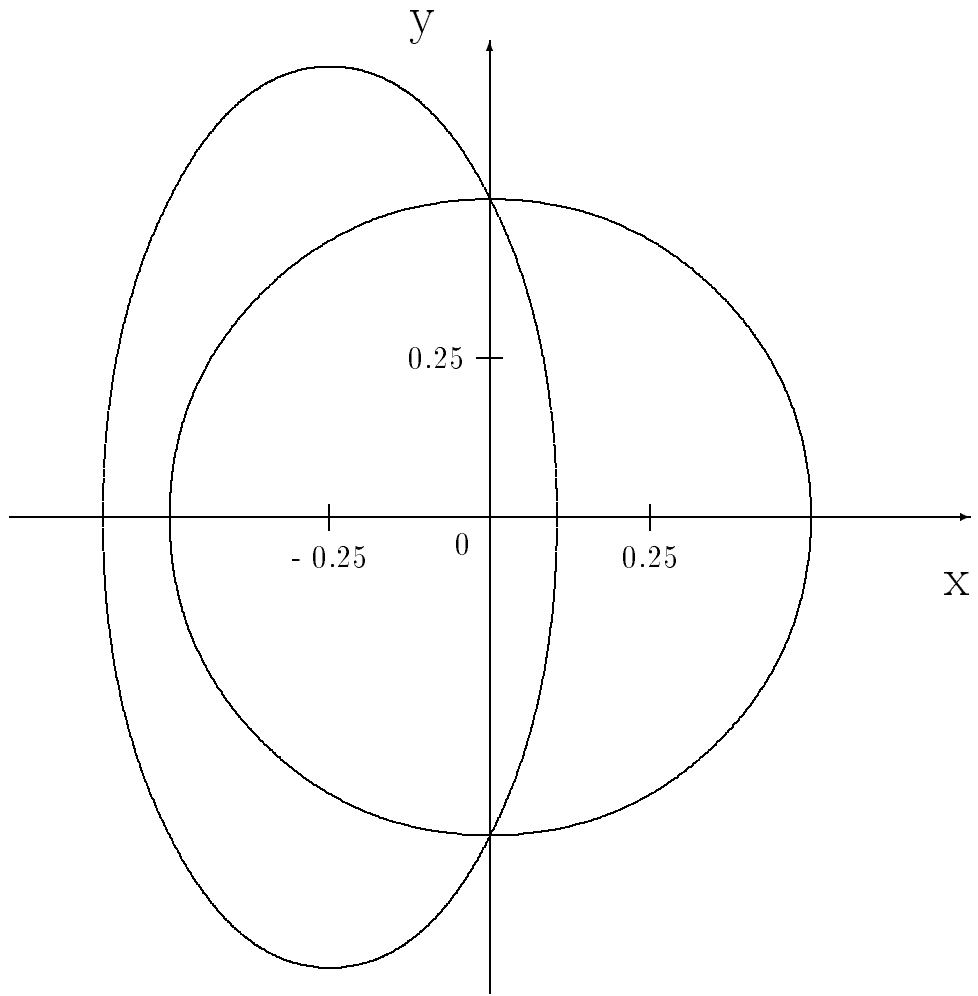
\begin{figure}
\caption{\label{xy}
Visualization of the field fluctuations before and after the measurement
for a measurement resolution of $\delta\!x=0.5$ and a measurement result
of $x_m=-0.5$. The contours shown mark the standard deviation of the 
Gaussian noise distributions. The circle represents the vacuum fluctuations.
After the measurement, the x-component is shifted by $x_m/2=-0.25$ and the
fluctuations in x are squeezed by a factor of $1/\sqrt{2}$. The fluctuations
in y are increased by a factor of $\sqrt{2}$ by the noise introduced in the 
measurement.}
\end{figure}

If the signal is in the vacuum state $\mid \mbox{vac.}\rangle$,
then the measurement probability is a Gaussian centered around
$x_m=0$ with a variance of $\delta\!x^2+1/4$,
\begin{equation}
\label{eq:vacprop}
P(x_m)= \frac{1}{\sqrt{2\pi (\delta\!x^2+1/4)}} 
\exp\left(-\frac{x_m^2}{2 (\delta\!x^2+1/4)}\right).
\end{equation}
The quantum state after the measurement is a squeezed state 
given by
\begin{equation}
\mid \phi_S(x_m)\rangle = \int d\!x_S\;
\left(\pi \frac{4\delta\!x^2}{1+4\delta\!x^2}\right)^{-\frac{1}{4}}
\exp\left(- \frac{1+4\delta\!x^2}{4\delta\!x^2}
           \left(x_S- \frac{x_m}{1+4\delta\!x^2}\right)^2\right) 
     \mid x_S \rangle.
\end{equation}
The quadrature component averages and variances of this state are 
\begin{eqnarray}
\langle \hat{x}_S \rangle_{x_m}&=& \frac{x_m}{1+4\delta\!x^2}
\nonumber \\[0.2cm]
\langle \hat{y}_S \rangle_{x_m}&=& 0
\nonumber \\[0.2cm]
\langle \hat{x}_S^2 \rangle_{x_m} - \langle \hat{x}_S \rangle_{x_m}^2
&=& \frac{\delta\!x^2}{1+4\delta\!x^2}
\nonumber \\[0.2cm]
\langle \hat{y}_S^2 \rangle_{x_m} - \langle \hat{y}_S \rangle_{x_m}^2
&=& \frac{1+4\delta\!x^2}{16\delta\!x^2}.
\end{eqnarray}
Examples of the phase space contours before and after the measurement
are shown in figure \ref{xy} for a measurement resolution of $\delta\!x=0.5$
and a measurement result of $x_m=-0.5$. Note that the final state is shifted
by only half the measurement result. 

The photon number expectation value after the measurement is given by
the expectation values of $\hat{x}_S^2$ and $\hat{y}_S^2$. It reads
\begin{eqnarray}
\label{eq:vacphoton}
\langle \hat{n}_S \rangle_{x_m} &=& \langle\hat{x}_S^2\rangle_{x_m}  
                          + \langle \hat{y}_S^2 \rangle_{x_m} - \frac{1}{2}
\nonumber \\
                        &=& \frac{1}{16 \delta\!x^2 (1+4\delta\!x^2)}
                            + \frac{x_m^2}{(1+4\delta\!x^2)^2}.
\end{eqnarray}
The dependence of the photon number expectation value 
$\langle \hat{n}_S \rangle_{x_m}$ after the measurement 
on the squared measurement result $x_m^2$ describes a correlation
between field component and photon number defined by
\begin{eqnarray}
C(x_m^2; \langle \hat{n}_S \rangle_{x_m}) &=&
\int
\left(\int d\!x_m\; x_m^2 \langle \hat{n}_S \rangle_{x_m} P(x_m)\right) 
- 
\left(\int d\!x_m\; x_m^2 P(x_m)\right)
\left(\int d\!x_m\; \langle \hat{n}_S \rangle_{x_m} P(x_m)\right).
\nonumber \\
\end{eqnarray}
According to equations (\ref{eq:vacprop}) and (\ref{eq:vacphoton}),
this correlation is equal to 
\begin{equation}
C(x_m^2; \langle \hat{n}_S \rangle_{x_m}) = \frac{1}{8}
\end{equation}
for measurements of the vacuum state. This result is independent of the
measurement resolution. In particular, it even applies to the low resolution
limit of $\delta\!x\to \infty$, which should leave the original
vacuum state nearly unchanged. It is therefore reasonable to conclude, that
this correlation is a fundamental property of the vacuum state, even though 
it involves nonzero photon numbers.
%=========================================================

\section{Correlations of photon number and fields in the operator formalism}
\label{sec:fundop}

Since the measurement readout $x_m$ represents information about
operator variable $\hat{x}_S$ of the system, it is possible to 
express the correlation 
$C(x_m^2; \langle \hat{n}_S \rangle_{x_m})$ in terms of 
operator expectation values of $\hat{x}_S$ and $\hat{n}_S$.
Equation (\ref{eq:av}) shows how the average over $x_m^2$
can be replaced by the operator expectation value 
$\langle \hat{x}_S^2 \rangle$. Likewise, the 
average over the product of $x_m^2$ and 
$\langle \hat{n}_S \rangle_{x_m}$ can be transformed into an
operator expression. The transformation reads
\begin{eqnarray}
\label{eq:trans}
\lefteqn{
\int d\!x_m\; x_m^2 \langle \hat{n}_S \rangle_{x_m} P(x_m) = }
\nonumber \\[0.2cm]
&=& \int  d\!x_S d\!x_S^\prime \left(\frac{(x_S+x_S^\prime)^2}{4} 
+ \delta\!x^2\right)
\langle \mbox{vac.}\mid x_S \rangle 
   \langle x_S \mid \hat{n}_S \mid x_S^\prime\rangle 
       \langle x_S^\prime \mid \mbox{vac.} \rangle 
\exp\left(-\frac{(x_S-x_S^\prime)^2}{8\delta\!x^2}\right)
\nonumber \\[0.2cm]
&=& \int  d\!x_m\; 
\left(\frac{1}{4}\langle \hat{x}_S^2\hat{n}_S + 2 \hat{x}_S\hat{n}_S\hat{x}_S 
+ \hat{n}_S\hat{x}_S^2 \rangle_{x_m} 
+ \delta\!x^2 \langle \hat{n}_S \rangle_{x_m}\right) P(x_m).
\end{eqnarray}
The average expectation value of photon number after the measurement
is given by
\begin{equation}
\langle \hat{n}_S \rangle_{\mbox{av.}} =  
\int  d\!x_m\; \langle\hat{n}_S \rangle_{x_m} P(x_m).
\end{equation}
Using the index $\mbox{av.}$ to denote averages over expectation values
after the measurement, the correlation 
$C(x_m^2; \langle \hat{n}_S \rangle_{x_m})$
may be expressed by the average final state expectation values as
\begin{equation}
C(x_m^2; \langle \hat{n}_S \rangle_{x_m}) =
\left(\frac{1}{4}\langle \hat{x}_S^2\hat{n}_S 
+ 2 \hat{x}_S \hat{n}_S \hat{x}_S
+ \hat{n}_S\hat{x}_S^2 \rangle_{\mbox{av.}}  
- \langle n_S \rangle_{\mbox{av.}} 
  \langle x_S^2 \rangle_{\mbox{av.}}\right).
\end{equation}
The correlation observed in the measurement is therefore given by a
particular ordered product of operators. The most significant feature
of this operator product is the $\hat{x}_S\hat{n}_S\hat{x}_S$-term,
in which the photon number operator $\hat{n}_S$ is sandwiched between 
the field operators $\hat{x}_S$. The expectation value of 
$\hat{x}_S\hat{n}_S\hat{x}_S$ of an eigenstate of $\hat{n}_S$ does not
factorize into the eigenvalue of $\hat{n}_S$ and the expectation value 
of $\hat{x}_S^2$, because the field operators $\hat{x}_S$ change the
original state into a state with different photon number statistics.  
According to the commutation relations,
\begin{equation}
\hat{x}_S\hat{n}_S\hat{x}_S = 
\frac{1}{2}(\hat{x}_S^2\hat{n}_S + \hat{n}_S\hat{x}_S^2) + \frac{1}{4}.
\end{equation}
Therefore, the expectation value of $\hat{x}_S\hat{n}_S\hat{x}_S$
of a photon number state is exactly $1/4$ higher than the product
of the eigenvalue of $\hat{n}_S$ and the expectation value of 
$\hat{x}_S^2$. The correlation $C(x_m^2; \langle \hat{n}_S \rangle_{x_m})$
may then be expressed by the final state expectation values as
\begin{equation}
C(x_m^2; \langle \hat{n}_S \rangle_{x_m}) =
\left(\frac{1}{2}\langle \hat{x}_S^2\hat{n}_S 
+ \hat{n}_S\hat{x}_S^2 \rangle_{\mbox{av.}}  
- \langle n_S \rangle_{\mbox{av.}} 
  \langle x_S^2 \rangle_{\mbox{av.}}\right)
+ \frac{1}{8}.
\end{equation}
Since the additional correlation of $1/8$ does not depend on the measurement
resolution $\delta\!x$, it should not be interpreted as a result of the 
measurement dynamics. Instead, the derivation above reveals that it 
originates from the 
operator ordering in the quantum mechanical expression for the correlation. 
Since it is the noncommutativity of operator variables which distinguishes 
quantum physics from classical physics, the contribution of $1/8$ is a 
nonclassical contribution to the correlation of photon number and fields. 
Specifically, it should be noted that the classical correlation of a well 
defined variable
with any other physical property is necessarily zero. Only the quantum 
mechanical properties of noncommutative variables allow nonzero
correlations of photon number and fields even if the field mode 
is in a photon number eigenstate. 
The operator transformation thus reveals that the correlation 
$C(x_m^2; \langle \hat{n}_S \rangle_{x_m})$ of $1/8$ found in measurements
of the vacuum state is a directly observable consequence of the nonclassical
operator order dependence of correlations between noncommuting variables.

%=========================================================

\section{Experimental realization: photon-field coincidence measurements}
\label{sec:ex}

The experimental setup required to measure the correlation between a
QND measurement of the quadrature component $\hat{x}_S$ and the
photon number after the measurement is shown in figure \ref{setup}.
It is essentially identical to the setups used in previous experiments 
\cite{Por89,Per94}. However, instead of measuring the x quadrature in the 
output fields, it is necessary to perform a photon number 
measurement on the signal branch. The output of this measurement
must then be correlated the output from the homodyne detection of the
meter branch. The homodyne detection of the meter simply 
converts a high intensity light field into a current $I_M(t)$,
while the signal readout produces discreet photon detection pulse.
These pulses can also be described by a detection current $I_S(t)$,
which should be related to the actual photon detection events by a
response function $R_S(\tau)$, such that
\begin{equation}
I_S(t) = \sum_i R_S(t-t_i),
\end{equation} 
where $t_i$ is the time of photon detection event $i$. 
According to the theoretical prediction discussed above, each
photon number detection event should be accompanied by an increase
of noise in the homodyne detection current of the meter.
However, the temporal overlap of the signal current $I_S(t)$
and the increased noise in the meter current $I_M(t)$ is
an important factor in the evaluation of the correlation.
Due to the frequency filtering employed, the meter mode corresponding
to a signal detection event is given by a filter function
with a width approximately equal to the inverse frequency resolution
of the filter. For a typical filter with a Lorentzian linewidth of
$2\gamma$, the mode of interest
would read
\begin{equation}
\label{eq:mode}
\hat{a}_i = 
\sqrt{\gamma} \int dt \exp\left(-\gamma\;|\;t-t_i\;|\right) \hat{a}(t).
\end{equation}
The actual meter readout should therefore be obtained by integrating
the current over a time of about $2/\gamma$. For practical reasons, it
seems most realistic to use a direct convolution of the meter
current $I_M$ and the signal current $I_S$, adjusting the response function
$R_S(\tau)$ to produce an electrical pulse of duration $2/\gamma$.
A measure of the correlation $C(x_m^2; \langle \hat{n}_S \rangle_{x_m})$
can then be obtained from the current correlation
\begin{equation}
\xi\;C(x_m^2; \langle \hat{n}_S \rangle_{x_m}) 
= \overline{(I_S I_M)^2} - \overline{I_S^2}\;\overline{I_M^2},
\end{equation}
where the factor $\xi$ denotes the efficiency of the measurement, as
determined by the match between the response function $R_S(\tau)$ and the
filter function given by equation (\ref{eq:mode}). Moreover, the
efficiency of the experimental setup may be reduced further by the
limited quantum efficiency of the detector. 

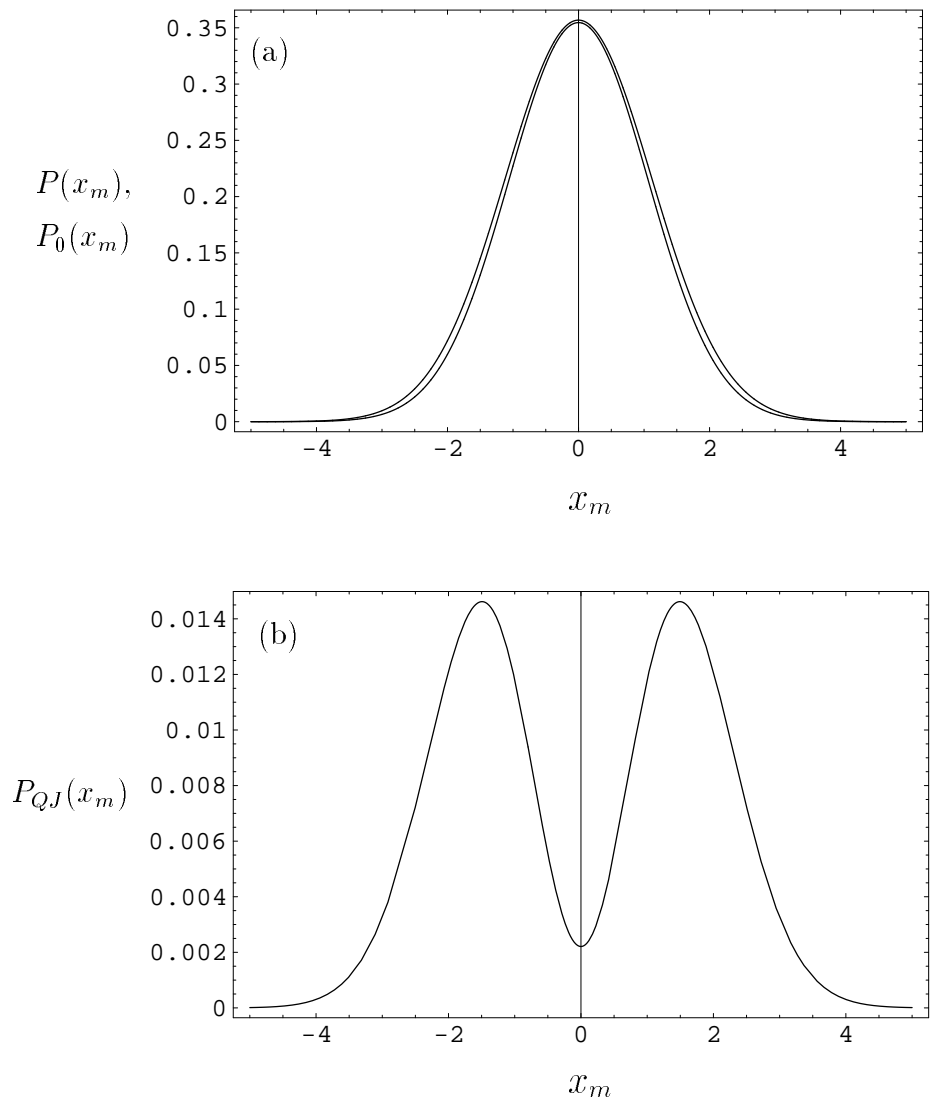
\begin{figure}
\caption{\label{qj}
Separation of the probability distribution $P(x_m)$ of the measurement 
result $x_m$ into a component $P_0(x_m)$ associated with no quantum jump
and a component $P_{QJ}(x_m)$ associated with a quantum jump to one or more
photons at a measurement resolution of $\delta\!x=1$.
(a) shows both $P(x_m)$ and $P_0(x_m)$, which are only slightly
different from each other. (b) shows the difference given by the quantum
jump contribution $P_{QJ}(x_m)$. The total probability of a quantum jump
at $\delta\!x=1$ is 5.72\%.}
\end{figure}

Fortunately, the requirement of efficiency for the 
experiment is not very restrictive, provided that the measurement resolution
is so low that only few photons are created. In that case, the total
noise average in the meter current $I_M$ is roughly equal to the noise
average in the absence of a photon detection event, which is very close to 
the shot noise limit of the homodyne detection. However, the fluctuations
of the time averaged currents within a time
interval of about $1/\gamma$ around a photon detection event in the 
signal branch correspond to the fluctuations of the measurement 
values $x_m$ for a quantum jump event from zero photons to one photon.
In particular, the measurement result $x_m(i)$ associated with a 
photon detection event at time $t_i$ is approximately given by
\begin{equation}
x_m (i) \approx C \int dt R(t-t_i) I_M(t),
\end{equation}
where $C$ is a scaling constant which maps the current fluctuations 
of a vacuum input field onto an $x_m$ variance of $\delta\!x^2$.
In the case of a photon detection event, however, the probability 
distribution over the measurement results $x_m (i)$ is given
by the difference between the total probability
distribution $P(x_m)$ and the part $P_0(x_m)$ of the probability distribution
associated with no photons in the signal,
\begin{eqnarray}
P_{QJ}(x_m) &=& P(x_m) - P_0(x_m)
\nonumber \\[.2cm]
&=& \langle \mbox{vac.}\mid \hat{P}_{\delta\!x}^2 \mid \mbox{vac.}\rangle
- \langle \mbox{vac.}\mid \hat{P}_{\delta\!x} \mid \mbox{vac.}\rangle^2
\nonumber \\
&=& \frac{1}{\sqrt{2\pi (\delta\!x^2+1/4)}} 
\exp\left(-\frac{x_m^2}{2 (\delta\!x^2+1/4)}\right) 
\;-\; \sqrt{\frac{32\delta\!x^2}{\pi(1+8\delta\!x^2)^2}} 
       \exp\left(-\frac{4}{1+8\delta\!x^2}x_m^2\right)
.
\end{eqnarray}
Figure \ref{qj} shows the results for a measurement resolution of 
$\delta\!x=1$, which is close to the experimentally realized resolution
reported in \cite{Per94}. There is only a slight difference in
$P(x_m)$ and $P_0(x_m)$, even though the total probability of a quantum 
jump to one or more photons obtained by integrating $P_{QJ}(x_m)$ 
is about 5.72\% . The peaks of the probability 
distribution are close to $\pm 2$, eight times higher than the fluctuation
of $\hat{x}_S$ in the vacuum. 
The measurement fluctuations corresponding to a photon detection event
are given by
\begin{equation}
\frac{\int d\!x_m\; x_m^2 P_{QJ}(x_m)}{\int d\!x_m P_{QJ}(x_m)}
= \frac{1}{4}+\delta\!x^2\left(2+\sqrt{1+\frac{1}{8\delta\!x^2}}\right)
\approx 3 \delta\!x^2. 
\end{equation}
For $\delta\!x\gg1$, this result is three times higher than the overall
average. For $\delta\!x=1$, the ratio between the fluctuation intensity 
of a detection event and the average fluctuation intensity of
$1/4+\delta\!x^2$ is still equal to 2.65. In other words, the fluctuations
of the measurement result $x_m$ nearly triple in the case of a quantum jump
event. The corresponding increase in the fluctuations of the homodyne 
detection current $I_M$ should be detectable even at low efficiencies
$\xi$. Moreover, it does not matter how many photon events go undetected, 
since the ratio has been determined relative to the overall average of
the meter fluctuations. It is thus possible to obtain experimental evidence
of the fundamental correlation of field component and photon number even
with a rather low overall efficiency of the detector setup.

%=========================================================

\section{Interpretation of the quantum jump statistics}
\label{sec:int}

What physical mechanism causes the quantum jump from the zero
photon vacuum to one or more photons? The relationship between 
the photon number operator and the quadrature components of the 
field is given by
\begin{equation}
\label{eq:ndef}
\hat{n}_S + \frac{1}{2} = \hat{x}_S^2 + \hat{y}_S^2.
\end{equation}
According to equation (\ref{eq:shift}) describing the measurement 
interaction, the change in photon number $\hat{n_S}$ should 
therefore be caused by the change in $\hat{y}_S$ caused by 
$\hat{y}_M$,
\begin{equation}
\hat{U}_{SM}^{-1}\;\hat{n}_S\;\hat{U}_{SM} = \hat{n}_S
- 2f\hat{y}_S\hat{y}_M + f^2\hat{y}_M^2.
\end{equation}
Thus the change in photon number does not depend explicitly on either
the measured quadrature $\hat{x}_S$ or the meter variable
$\hat{x}_M$. Nevertheless, the meter readout shows a strong correlation 
with the quantum jump events. In particular, the probability distribution 
of meter readout results $x_m$ for a quantum jump to one or more photons
shown in figure \ref{qj} has peaks at values far outside the range given
by the variance of the vacuum fluctuations of $\hat{x}_S$. 

Moreover, the correlation between readout and photon number after the 
measurement does not disappear in the limit of low resolution
($\delta\!x\to\infty$). Rather, it appears to be a fundamental
property of the vacuum state even before the measurement. This is confirmed 
by the operator formalism, which identifies the source of the correlation
as the expectation value $\langle\hat{x}_S\hat{n}_S\hat{x}_S\rangle$. 
This expectation
value is equal to $1/4$ in the vacuum, even though the photon number is 
zero. Since the operator formalism does not allow an identification of the
operator with the eigenvalue unless it acts directly on the eigenstate,
it is possible to find nonzero correlations even if the system is in an 
eigenstate of one of the correlated variables. In particular, the action
of the operator $\hat{x}_S$ on the vacuum state is given by
\begin{equation}
\hat{x}_S\mid\mbox{vac.}\rangle = \frac{1}{2}\mid n_s=1 \rangle,
\end{equation}
so the operator $\hat{x}_S$ which should only determine the 
statistical properties of the state with regard to the quadrature component
$x_S$ changes the vacuum state into the one photon state. The 
application of operators thus causes fluctuations in a variable even when the 
eigenvalue of that variable is well defined. 

The nature of this fluctuation might be clarified by a comparison
of the nonclassical correlation obtained for fields and photon number 
in the vacuum with the results of quantum tomography by homodyne 
detection\cite{Vog89,Smi93}. In such 
measurements, the photon number is never obtained. Rather, the complete
Wigner distribution $W(x_S,y_S)$ can be reconstructed from the results. 
It is therefore possible to deduce correlations between the field 
components and the field intensity defined by $I = x_S^2+y_S^2$, 
which is the classical equivalent of
equation (\ref{eq:ndef}). For the vacuum, the Wigner function reads
\begin{equation}
\int d\!x_S d\!y_S\; x_S^4 W_0(x_S,y_S) - (\int d\!x_S d\!y_S\; x_S^2 W_0(x_S,y_S))^2
= 1/8.
\end{equation} 
The correlation of $I$ and $x_S^2$ is given by
\begin{eqnarray}
\label{eq:wigcor}
\lefteqn{C(x_S^2; I) =}
\nonumber \\ &&
\int
\left(\int d\!x_S d\!y_S \; x_S^2\;I \; W_0(x_S,y_S)\right) 
-
\left(\int d\!x_S d\!y_S \; x_S^2 \; W_0(x_S,y_S)\right)
\left(\int d\!x_S d\!x_S \; I \; W_0(x_S,y_S)\right) 
\nonumber \\ 
&=& C(x_m^2; \langle n_S \rangle_{x_m})=\frac{1}{8}.
\end{eqnarray}
Thus, the correlation between $I=x_S^2+y_S^2$ and $x_S^2$ described by
the Wigner distribution is also equal to $1/8$. In fact, the ``intensity
fluctuations'' of the Wigner function can be traced to the same operator 
properties that give rise to the correlations between the field measurement
result and the induced photon number. For arbitrary signal fields, the
correlation between the squared measurement result and the photon number
after the measurement can therefore be derived by integrating over the 
Wigner function of the signal field after the measurement interaction
according to equation (\ref{eq:wigcor}).

Of course the ``intensity fluctuations'' of the Wigner function 
cannot be observed directly, since any phase insensitive determination
of photon number will reveal the well defined result of zero photons in
the vacuum. Nevertheless even a low resolution measurement of the quadrature 
component $\hat{x}_S$ which leaves the vacuum state nearly unchanged
reveals a correlation of $\hat{x}_S^2$ and $n_S$ which corresponds to the
assumption that the measured quadrature $\hat{x}_S$ contributes to a
fluctuating vacuum energy. The quantum jump itself appears to draw its
energy not from the external influence of the measurement interaction, but
from the fluctuating energy contribution $\hat{x}_S^2$. These energy 
fluctuations could be interpreted as virtual or hidden fluctuations
existing only potentially until the energy uncertainty of the measurement
interaction removes the constraints imposed by quantization and energy
conservation. 
In particular, energy conservation does require that the energy for the 
quantum jump is provided by the optical parametric amplification process.
Certainly the {\it average} energy is supplied by the pump beam. However,
the energy content of the pump beam and the meter beam cannot be defined 
due to the uncertainty principle. The pump must be coherent and the 
measurement of the meter field component $\hat{x}_M$ prevents all energy
measurements in that field. If it is accepted that quantum mechanical 
reality is somehow conditioned by the circumstances of the measurement, 
it can be argued that the reality of quantized photon number only exists if the
energy exchange of the system with the environment is controlled on the
level of single quanta. Otherwise, it is entirely possible that the vacuum
energy might not be zero as suggested by the photon number eigenvalue,
but might fluctuate according to the statistics suggested by the Wigner 
function. 

Even though it may appear to be highly unorthodox at first, this 
``relaxation'' of quantization rules actually corresponds
to the noncommutativity of the operators, and may help explain the seemingly
nonlocal properties of entanglement associated with the 
famous EPR paradox \cite{EPR}. The definition of elements of reality
given by EPR reads
``{\it If, without in any way disturbing a system, we can predict
with certainty (i.e., with probability equal to unity) the value of a 
physical quantity, then there exists an element of physical reality
corresponding to this physical quantity.}''
This definition of elements of reality assumes that the eigenvalues of
quantum states are real even if they are not confirmed in future
measurements. In particular, the photon number of the vacuum would 
be considered as a real number, not an operator, so   
the operator correlation $\langle\hat{x}_S\hat{n}_S\hat{x}_S\rangle$ 
should not have any physical meaning.
However, the nonzero correlation of fields and
photon number in the vacuum observed in the QND measurement discussed
above suggests that {\it even the possibility
of predicting the value of a physical quantity with certainty only 
defines an element of reality if this value is directly observed
in a measurement}. Based on this conclusion, there is no need to
assume any ``spooky action at a distance'', or physical nonlocality,
in order to explain Bell's inequalities \cite{Bel64}. Instead, it is 
sufficient to point out that knowledge of the wavefunction does not provide
knowledge of the type of measurement that will be performed.
In the case of spin-1/2 systems, the quantized values of spin 
components are not a property inherent in the spin system, but a 
property of the measurement
actually performed. To assume that spins are quantized even without 
a measurement does not correspond to the implications of the operator 
formalism, since it is not correct to replace operators with their 
eigenvalues. 

In the same manner, the correlation discussed in this paper would be 
paradoxical if one regarded the photon number eigenvalue of zero in the
vacuum state as an element of reality independent of the measurement
actually performed. One would then be forced to construct mysterious
forces changing the photon number in response to the measurement result.
However, the operator formalism suggests no such hidden forces. Instead,
the reality of photon number quantization depends on the operator ordering
and thus proofs to be rather fragile.

%=========================================================

\section{Summary and conclusions}
\label{sec:concl}

The change in photon number induced by a quantum nondemolition measurement
of a quadrature component of the vacuum is strongly correlated with
the measurement result. An experimental determination of this correlation
is possible using optical parametric amplification in a setup similar 
to previously realized QND measurements of quadrature components
\cite{Por89,Per94}. The observed correlation corresponds to a fundamental
property of the operator formalism which allows nonvanishing correlations
between noncommuting variables even if the system is in an eigenstate
of one of the variables.

The quantum jump probability reflects the properties of intensity 
fluctuations corresponding to the vacuum fluctuations of the field
components. The total correlation of fields and photon number
therefore reproduces the result that would be expected if there was 
no quantization. It seems that quantum jumps are a mechanism by 
which the correspondence between quantum mechanics and classical physics
is ensured. The quantum jump correlation observable in the
experimental situation discussed above thus provides a link
between the discrete nature of quantized information and the continuous
nature of classical signals. Finite resolution QND measurements 
could therefore provide a more detailed understanding of the nonclassical 
properties of quantum information in the light field.

\section*{Acknowledgements}
One of us (HFH) would like to acknowledge support from the Japanese 
Society for the Promotion of Science, JSPS.
%=========================================================

%========================================================================

\begin{figure}
\caption{\label{setup}
Schematic illustration of the measurement setup for a back action evasion 
quantum nondemolition measurement of a quadrature component using optical 
parametric amplifiers (OPAs). Note that the reflectivity of the beam 
splitters depends on the amplification achieved in the parametric 
downconversion process. The coupling factor for the measurement is given 
by $f=(a^2-1)/a$.
}
\end{figure}

%=========================================================

\end{document}